\documentclass[preprint,prl,aps,floats]{revtex4}
\usepackage{graphicx}
\begin{document}
\title{Generalized Thermostatistics and Bose-Einstein Condensation}
\author{H. G. Miller$^1$\thanks{E-Mail:   hmiller@maple.up.ac.za}, F. C. Khanna$^{2,3}$, R.Teshima$^2$, A.R. Plastino$^1$ and A. Plastino$^4$}
\affiliation{
         $^1$Department of Physics, University of Pretoria, Pretoria 0002, South Africa\\
    $^2$ Theoretical Physics Institute, Department of Physics,
       University of Alberta, Edmonton, Alberta, Canada T6G~2J1 \\
     $^3$TRIUMF, 4004 Wesbrook Mall, Vancouver, British Columbia,
        Canada V6T~2A3\\
    $^4$ La Plata Physics Institute \\ Exact Sciences Faculty-National
    University La Plata and CONICET\\C. C. 727, 1900 La Plata,
Argentina}
\begin{abstract}
Analytical expressions for Bose-Einstein condensation of an ideal
Bose gas analyzed within the strictures of non-extensive,
generalized thermostatistics are here obtained.
\end{abstract}
\maketitle

Nonextensive thermostatistics (NEXT) \cite{euro,TG04,T88,T01} has
been utilized with success in connection with a number of problems
both in the classical \cite{F05,TGS05,B01,B04,T00,TTL02,PLR04} and
the quantum regimes \cite{C96,TBD98,UMK01,TMT03,PPMU04}. NEXT is
thought to be relevant for the study (among others) of: systems
described by non linear Fokker-Planck equations \cite{F05};
systems with a scale-invariant occupancy of phase space
\cite{TGS05}; non equilibrium scenarios involving temperature
fluctuations \cite{B01,B04}; systems exhibiting weak chaos
\cite{T00,TTL02}; and systems with interactions of long range
relative to the system's size \cite{PLR04}. This list is far from
complete. For reviews on the applications of the NEXT formalism
see \cite{euro,T01,TG04}.

  The applications of non-extensive thermostatistics to the
classical domain are much more developed than applications to
quantum mechanical problems. The (comparatively) slow progress
made in applying NEXT ideas to quantum systems may be due, at
least in part, to the considerable difficulty in obtaining
analytical results concerning basic (statistical) aspects of
quantum many-body physics. The aim of this letter is to report on
one such result, in connection with the phenomenon of
Bose-Einstein condensation (BEC). The recent successful
experimental realization of BEC in atomic gases
\cite{a+95,D+95,CSH97} has led to considerable theoretical and
experimental activity \cite{DGPS99,PW98,BK91,ET03,C96}. In point
of fact, some of the theoretical efforts have focused on the role
of the statistics on BEC \cite{C96} particularly when changing
from the extensive one based upon the Boltzmann-Gibbs (BG)
logarithmic entropy, to the generalized
version proposed in Ref. \cite{T88}.

Within the standard Boltzmann-Gibbs' thermostatistical formalism,
an ideal Bose-Einstein gas is the simplest exactly solvable
continuous system that undergoes a phase transition. Generalized
non-extensive statistics is characterized by a non-logarithmic
entropy $S_q$ that contains a free parameter $q$ called the
non-extensivity index (for $q \rightarrow 1$, one has $S_q
\rightarrow$ Boltzmann's $S$).  Application of Jaynes' MaxEnt
methodology \cite{katz} within a NEXT-context yields power-law
probability distributions \cite{PP93}. Previous $S_q-$studies of
BEC have only been undertaken in restricted regions \cite{C96} in
the vicinity of the $q\rightarrow 1$ limit where BG-statistics is
recovered. In this communication we considerably amplify this
rather restricted scope by generating NEXT-exact analytic
expressions, and then determine the all-important $q-$values for
which BEC may occur.

Consider an ideal gas of particles of mass $m$ in $d$ dimensions
which obey Bose-Einstein statistics. In the BG case the average
number of particles in the grand canonical ensemble is given by
\begin{eqnarray}
N^{BG}_d &=& c_d\int^\infty_0
\frac{p^{d-1}}{z^{-1}e^{\beta\epsilon}-1}dp \\
&=&c_d\left( \frac{1}{2} \right)^d
\left(\frac{2m}{\beta} \right)^{\frac{d}{2}}
\int^\infty_0\frac{x^{\frac
{d}{2}-1}}{z^{-1}e^x -1} dx \\
 &=& c_d \left( \frac{1}{2} \right)^d
\left( \frac{2m}{\beta} \right)^{\frac{d}{2}}
\Gamma\left(\frac{d}{2} \right)
g_{\frac{d}{2}}(z)
\end{eqnarray}
where $z=e^{\beta\mu}$, $\epsilon=\frac{p^2}{2m}$,
$\beta=\frac{1}{T}$,  $\mu$ is the chemical potential,
$g_\nu(z)=\frac{1}{\Gamma(\frac{d}{2})}
\int^\infty_0\frac{x^{\frac{d}{2}-1}}{z^{-1}e^x -1} dx
=\Sigma_{j=1}^\infty \frac{z^j}{j^\nu}$ is the well known
Bose-Einstein integral \cite{P84} and $c_d$ is a constant which
depends on the intergration over the other phase space variable of
dimension d. Since occupation numbers cannot be negative or
infinite, $\mu < 0$ and $0\leq z < 1$. For the limiting value $z
\rightarrow 1$, $g_\nu (z) \rightarrow \zeta(\nu)$ where $\zeta
(x)$ is the Riemann zeta function and hence only for d=3 does
Bose-Einstein condensation occur.

In the case where NEXT is employed, the average number of
particles is given by
\begin{eqnarray}
N^{TS}_d &=& c_d\int^\infty_0 \frac{p^{d-1}}{\left[ 1+(q-1)\beta
(\frac{p^2}{2m}-\mu)\right]^\frac{1}{q-1}-1} dp \\
         &=&c_d\int^\infty_0\frac{p^{d-1}}{(a+bp^2)^\frac{1}{q-1} -1} dp
\label{ts}
\end{eqnarray}
where
\[
a=1-(q-1)\mu\beta
\]
and
\[
b =\frac{(q-1)\beta}{2m}.
\]
Here we have used the power-law
 distribution function for
bosons \cite{PPMU04}.  For $\mu < 0$ and $q >1$ equation(\ref{ts})
may be written as
\begin{eqnarray}
N^{TS}_d &=&
c_d\int^\infty_0\Sigma_{n=0}^\infty\frac{p^{d-1}}{(a+bp^2)^\frac{n+1}{q-1} }
dp\\
&=&
c_d\Sigma_{n=0}^\infty\frac{(\frac{a}{b})^\frac{d}{2}}{a^\frac{n+1}{q-1}}
\int^\infty_0\frac{x^{d-1}}{(1+x^2)^\frac{n+1}{q-1} } dx \\
 &=&
c_d\Sigma_{n=0}^\infty\frac{\left(\frac{2m(1-(q-1)\mu\beta)}
{(q-1)\beta }\right)^\frac{d}{2}}
{(1-(q-1)\mu\beta)^\frac{n+1}{q-1}}
\frac{\frac{1}{2}\Gamma(\frac{d}{2})
\Gamma(\frac{n+1}{q-1}-\frac{d}{2})}
{\Gamma(\frac{n+1}{q-1})}
\end{eqnarray}
provided $0<d<2\frac{n+1}{q-1}$. For $\mu \rightarrow 0$
\begin{equation}
N^{TS}_d \rightarrow
c_d \left(\frac{2m}{(q-1)\beta} \right)^\frac{d}{2}\frac{1}{2}
\Gamma
\left( \frac{d}{2} \right)
\Sigma_{n=0}^\infty
\frac{\Gamma(\frac{n+1}{q-1}-\frac{d}{2})}{\Gamma(\frac{n+1}{q-1})}.
\label{mulim}
\end{equation}
In the limit $\mu \rightarrow 0$ the sum in eq(\ref{mulim}) is only convergent
for d=3 since for large z, $\frac{\Gamma(z-a)}{\Gamma(z)} \rightarrow \frac{1}{z^a}$.
 Furthermore in this case $q<5/3$.

 Hence, utilizing NEXT, rather than BG, does not permit BEC to occur in systems with dimensionality less than 3.
 In 3 dimensions
  BEC ocurs both in the BG and NEXT treatments, in the latter case only for $1<q<5/3$.

 \vspace{8mm}
 HGM acknowledges the hospitality of the Theoretical Physics Institute of the  Department of Physics at the
       University of Alberta and
the Physics Department of SUNY at Fredonia where part of this work was undertaken.


\end{document}